\documentclass{PoS}

\usepackage{epsfig}
\usepackage{latexsym}
\usepackage{footnote}

\def\spose#1{\hbox to 0pt{#1\hss}}
\def\ltapprox{\mathrel{\spose{\lower 3pt\hbox{$\mathchar"218$}}
 \raise 2.0pt\hbox{$\mathchar"13C$}}}
\def\gtapprox{\mathrel{\spose{\lower 3pt\hbox{$\mathchar"218$}}
 \raise 2.0pt\hbox{$\mathchar"13E$}}}

\title{Crossing the Gribov horizon: \\[2mm]
       an unconventional study of geometric properties \\[2mm]
       of gauge-configuration space in Landau gauge}

\ShortTitle{Crossing the Gribov horizon} 

\author{\speaker{Attilio Cucchieri}\thanks{We acknowledge
partial support from FAPESP ({\bf grant \# 2009/50180-0}) and from
CNPq. We would like to acknowledge computing time provided on the
Blue Gene/P supercomputer supported by the Research Computing
Support Group (Rice University) and Laborat\'orio de Computa\c c\~ao
Cient\'\i fica Avan\c cada (Universidade de S\~ao Paulo).}\\
        Instituto de F\'\i sica de S\~ao Carlos, Universidade de S\~ao Paulo,
        Caixa Postal 369, 13560-970 S\~ao Carlos, SP, Brazil\\
        E-mail: \email{attilio@ifsc.usp.br}}

\author{Tereza Mendes\\
        Instituto de F\'\i sica de S\~ao Carlos, Universidade de S\~ao Paulo,
        Caixa Postal 369, 13560-970 S\~ao Carlos, SP, Brazil\\
        E-mail: \email{mendes@ifsc.usp.br}}

\abstract{
We prove a lower bound for the smallest nonzero eigenvalue
of the Landau-gauge Faddeev-Popov matrix in Yang-Mills theories. 
The bound is written in terms of the smallest nonzero momentum on the
lattice and of a parameter
characterizing the geometry of the first Gribov region.
This allows a simple and intuitive description of the
infinite-volume limit in the ghost sector.
In particular, we show how nonperturbative effects may be
quantified by the rate at which typical thermalized and
gauge-fixed configurations approach the Gribov horizon.
Our analytic results are verified numerically in the SU(2) case
through an informal, free and easy, approach. This analysis provides
the first concrete explanation of why the so-called scaling solution
of the Dyson-Schwinger equations is not observed in lattice studies.}

\FullConference{31st International Symposium on Lattice Field Theory - LATTICE 2013\\
		July 29 - August 3, 2013\\
		Mainz, Germany}

\begin{document}


\section{Infinite-Volume Limit and the Boundary of the First Gribov
         Region $\Omega$}

Since the original work by Gribov \cite{Gribov:1977wm},
innumerous numerical and analytic studies (see, for example,
the reviews \cite{Cucchieri:2010xr,Vandersickel:2012tz}) have
focused on the infrared (IR) behavior of Yang-Mills Green's
functions in minimal-Landau gauge and on its connection to color
confinement \cite{Greensite:2009zz}. On the lattice, a detailed
description of the IR sector of Yang-Mills theories requires an
extrapolation to infinite volume. In minimal-Landau gauge, this
extrapolation should be governed by a (widely accepted)
axiom\footnote{This axiom is explained by considering the
interplay among the volume of configuration space, the
Boltzmann weight associated to the gauge configurations
and the step function used to constrain the functional
integration to the region $\Omega$.} stating that
\vskip 4mm
\begin{center}
\parbox{5.4in}{
\noindent
{\em At very large volumes, the functional integration gets
     concentrated on the boundary $\partial\Omega$ of the first 
     Gribov region $\Omega$ [defined by transverse gauge
     configurations with all nonnegative eigenvalues of the
     Faddeev-Popov (FP) matrix ${\cal M}$]. 
}}
\end{center}
\vskip 0.4mm 
\noindent
Thus, the functional integration should be strongly dominated at
very large volumes by configurations belonging to a thin layer
close to $\partial \Omega$, i.e.\ typical configurations should be
characterized by very small values for the smallest nonzero
eigenvalue $\lambda_1$ of ${\cal M}$. Indeed, numerical studies
show that $\lambda_1$ goes to zero as the lattice volume increases.

\vskip 3mm
In Ref.\ \cite{Cucchieri:2008fc} we have introduced the following
inequalities for the ghost propagator $G(p)$ in momentum space 
\begin{equation}
\frac{1}{N_c^2 - 1} \,  \frac{1}{\lambda_1} \, \sum_b \,
  | {\widetilde \psi_1(b,p)} |^2 \; \leq \; G(p) 
      \; \leq \; \frac{1}{\lambda_1} \; ,
\label{eq:lowerb}
\label{eq:upperb}
\end{equation}
where ${\widetilde \psi_1(b,p)}$ is the (Fourier-transformed)
eigenvector of the FP matrix ${\cal M}$ corresponding to the
eigenvalue $\lambda_1$ and $b = 1, 2, \ldots, N_c^2 - 1$ is
a color index running over the $N_c^2 - 1$ generators of the
SU($N_c$) gauge group. Note that, on the lattice and for large lattice
volumes, the smallest nonzero momentum $p_{min}$ is of the
order of $1/L$, where $L$ is the lattice size. Thus, if
$\lambda_1$ behaves as $\, L^{-2-\alpha} \,$ in the infinite-volume
limit, the inequality $\alpha > 0$ is a necessary condition
to obtain an IR-enhanced ghost propagator $G(p_{min}) \sim 1/
p_{min}^{2+2\kappa}$ (with $\kappa > 0$), as predicted in the
Gribov-Zwanziger confinement scenario \cite{Vandersickel:2012tz}
and by the scaling solution \cite{von Smekal:1997vx} of the
Dyson-Schwinger equations (DSEs) for gluon and ghost propagators.
At the same time, if the quantity $ \, | {\widetilde \psi_1(b,
p_{min})} |^2 \, $ behaves as $L^{-\gamma}$ at large $L$, the
inequality $\alpha - \gamma > 0$ is a sufficient condition to have
$\kappa > 0$. Therefore, in order to describe the extrapolation
to infinite volume in the ghost sector, we can re-formulate the
above axiom and say that
\vskip 4mm
\begin{center}
\parbox{5.4in}{
\noindent
{\em The key point seems to be the rate}\protect\footnote{At the
     same time, one needs
     a good projection of the eigenvector $\psi_1(b,x)$ on the
     plane waves corresponding to the momentum $p_{min}.$, i.e.\ the
     exponent $\gamma$ should not be too large.}
{\em at which $\lambda_1$
     goes to zero, which, in turn, should be related to the
     rate at which a thermalized and gauge-fixed configuration
     approaches $\partial \Omega$.}}
\end{center}
\vskip 0.4mm
\noindent
This is, however, just a qualitative statement. Indeed, in order
to make the above axiom quantitative, we need to relate the
eigenvalue $\lambda_1$ to the geometry of the Gribov region
$\Omega \,$. A first step in this direction was taken
in Ref.\ \cite{Cucchieri:2013nja}. There, we proved a lower
bound for $\lambda_1$ [see Eq.\ (\ref{eq:ineqnew2}) below]
that relates this eigenvalue to the distance of the gauge
configuration $A \in \Omega$ from the boundary $\partial
\Omega$. As a consequence, we were able to provide the first concrete
explanation of why the scaling solution of the DSEs is not 
observed in lattice studies.
The main results of Ref.\ \cite{Cucchieri:2013nja} are presented below.


\section{Lower bound for $\lambda_1$}

The (lattice) Landau gauge is usually imposed by minimizing the
functional\footnote{Here, we indicate with $e_{\mu}$ a unit
vector in the positive $\mu$ direction and with $a$ the lattice
spacing.}
\begin{equation}
{\cal E}[U;\omega] \,=\, - Tr \, \sum_{x, \mu} \,
\omega(x)\;U_{\mu}(x)\;\omega^{\dagger}(x+ a \,e_{\mu})
\label{eq:calE}
\end{equation}
with respect to the lattice gauge transformations $\,\omega(x) \in $
SU($N_c$). This defines the first Gribov region\footnote{For the
gauge field $A$ we consider the usual (unimproved) lattice
definition (see for example Ref.\ \cite{Cucchieri:2010xr}).}
\begin{equation}
\Omega \, \equiv \, \left\{\, U: \, \partial \cdot A = 0 \mbox{,}
\;\quad
{\cal M} \, = \, -D\cdot\partial \,\geq\,0\, \right\} \; ,
\label{eq:defOmega}
\end{equation}
where $D^{bc}(x,y)[A]$ is the covariant derivative and
${\cal M}(b,x;c,y)[A]$ is the FP matrix. One can show
\cite{Dell'Antonio:1991xt} that all gauge orbits intersect $\Omega$
and that this region is characterized by the following three
properties \cite{Zwanziger:1982na} (see also \cite{Vandersickel:2012tz,
Cucchieri:2013nja}):
{\bf 1}) the trivial vacuum $A_{\mu} = 0$ belongs to $\Omega$;
{\bf 2}) the region $\Omega$ is convex;
{\bf 3}) the region $\Omega$ is bounded in every direction.

From the definition of ${\cal M}[A]$ [see Eq.\ (\ref{eq:defOmega})]
it is clear that the FP matrix has a trivial null eigenvalue,
corresponding to constant vectors. Then, if we indicate with
$\lambda_1\left[ \, {\cal M}[A] \, \right]$ the smallest nonzero
eigenvalue of ${\cal M}[A]$, we can introduce the definition
\begin{equation}
\lambda_1\left[ \, {\cal M}[A] \, \right] \, = \,
\min_{\chi} \, \left( \chi \, , \left[ \, {\cal M}[A] \, \right] \,
\chi \right) \; ,
\label{eq:lambda1}
\end{equation}
where $\chi$ are non-constant vectors such that $( \chi \, ,
\chi) = 1$. Also, by noticing that the operators $D[A]$, ${\cal M}[A]
= -\partial^2 + {\cal K}[A]$ and ${\cal K}[A]$ are linear in the
gauge field $A$, one can write
\begin{equation}
{\cal M}[\rho A] \, = \, - \partial^2 + {\cal K}[\rho A]
  \, = \, (1 - \rho) \, (- \partial^2) \, + \, \rho
                       \, {\cal M}[A] \; .
\label{eq:Mlinear}
\end{equation}
At the same time, for $A \in \Omega$, $\rho \in [0, 1]$ and using the
first and second properties above, we have\footnote{This follows
immediately if we write $\rho A$ as the convex combination $(1-\rho)
A_1 + \rho A_2$, with $A_1 = 0$ and $A_2 = A$.} that $\rho A \in
\Omega$. This result applies in particular to configurations $A'$
belonging to the boundary $\partial \Omega$ of $\Omega$. Thus, by
using the definition (\ref{eq:lambda1}), Eq.\ (\ref{eq:Mlinear}) in
the case $A' \in \partial \Omega$ and the concavity of the minimum 
function \cite{matrixalgebra}, i.e.\ $ \, \min_{\chi} \left( \chi ,
\left[ M_1 + M_2 \right] \chi \right) \, \geq \, \min_{\chi}
\left( \chi , M_1 \chi \right) +  \min_{\chi} \left( \chi ,
M_2 \chi \right) \, $ (where $M_1$and $M_2$ are
two generic square matrices), we obtain \cite{Cucchieri:2013nja}
\begin{eqnarray}
\lambda_1\left[ \, {\cal M}[\rho A'] \, \right] & = & 
\lambda_1\left[(1 - \rho) \, (- \partial^2)
      \, + \, \rho \, {\cal M}[A']\right] \\[2mm]
& = & \min_{\chi} \,
      \left( \chi \, , \left[ \, (1 - \rho) \, (- \partial^2)
 \, + \, \rho \, {\cal M}[A'] \, \right] \, \chi \right) \\[2mm]
& \geq & (1 - \rho) \min_{\chi}
         \left( \chi , (- \partial^2) \, \chi \right) \, + \,
          \rho \min_{\chi} \, \left( \chi \, ,
             {\cal M}[A'] \, \chi \right) \; .
\end{eqnarray}
Since $A' \in \partial \Omega$, i.e.\ the smallest
non-trivial eigenvalue of the FP matrix ${\cal M}[A']$ is
null, and since the smallest non-trivial eigenvalue of (minus) the
Laplacian $ - \partial^2$ is $p^2_{min}$, we find
\begin{equation}
\lambda_1\left[ \, {\cal M}[\rho A'] \, \right] \, \geq \,
  (1 - \rho) \, p^2_{min} \; .
\label{eq:ineqnew}
\end{equation}
Therefore, as the lattice size $L$ goes to infinity, the eigenvalue
$\lambda_1\left[ \, {\cal M}[\rho A'] \, \right]$ cannot go to zero
faster than $(1 - \rho) \, p^2_{min}$. In particular, since $p^2_{min}
\sim 1/L^2$ for large lattice size $L$, we have that $\lambda_1$ behaves
as $\, L^{-2-\alpha} \,$ in the same limit, with $\alpha > 0$, only if 
$1 - \rho$ goes to zero at least as fast as $L^{-\alpha}$. Let us stress
that this result applies to any Gribov copy belonging to $\Omega$.

With $\rho A' = A$, the above inequality (\ref{eq:ineqnew}) may also be
written as
\begin{equation}
\lambda_1\left[ \, {\cal M}[A] \, \right]
\;\geq\; [1 - \rho] \, p^2_{min} \; .
\label{eq:ineqnew2}
\end{equation}
Note that, in the Abelian case, one has ${\cal M} = - \partial^2$ and
$\lambda_1 = p^2_{min}$, i.e.\ non-Abelian effects are included in
the factor $(1 - \rho)$. At the same time, the
quantity $1 - \rho\leq 1$ measures the distance (see Ref.\
\cite{Cucchieri:2013nja} for details) of a configuration $A \in \Omega$
from the boundary $\partial \Omega$ (in such a way that $A' = \rho^{-1}
A \in \partial \Omega$). Thus, the new bound (\ref{eq:ineqnew2})
suggests all non-perturbative features of a minimal-Landau-gauge
configuration $A \in \Omega$ to be related to its normalized distance
$\rho$ from the ``origin'' $A=0$ [or, equivalently, to its normalized
distance $1-\rho$ from the boundary $\partial \Omega$]. One should
also stress that the above inequality becomes an equality if and only if the
eigenvectors corresponding to the smallest nonzero eigenvalues of
${\cal M}[A]$ and $-\partial^2$ coincide.

As a consequence of the result (\ref{eq:ineqnew2}), we can also find
several new bounds. In particular, using the upper bound in 
Eq.\ (\ref{eq:upperb}) and the definition of the
Gribov ghost form-factor $\sigma(p)$, we have
\begin{equation}
\frac{1}{p^2_{min}} \, \frac{1}{1 - \sigma(p_{min})} 
\; \equiv \; G(p_{min}) \; \leq \;
       \frac{1}{[1 - \rho] \, p^2_{min}}
\label{eq:ineqG}
\end{equation}
and therefore $ \, \sigma(p_{min}) \, \leq \, \rho$.
This result is a stronger version of the so-called no-pole
condition\footnote{See, for example, \cite{Cucchieri:2012cb} and
references therein.} $ \, \sigma(p) \, \leq \, 1 \, $ (for
$p^2 > 0$), used to impose the restriction of the physical
configuration space to the region $\Omega$. Similarly, for the
horizon function $H$, defined in Eq.\ (3.10) of
\cite{Zwanziger:1991ac}, one can prove that \cite{Cucchieri:2013nja}
\begin{equation}
\frac{H}{d V (N_c^2 - 1)} \; \equiv \; 
   h \; \leq \; \rho \; . 
\end{equation}


\begin{table}[t]
\begin{center}
\begin{tabular}{|c|c|c|c|c|c|} \hline
$N$ &  $\max(n)$ & $\min(n)$ & $\langle n \rangle$
   & $R_{\rm before}/1000$  & $R_{\rm after}/1000$ \\ \hline
16    &   30         &         6          & 17.2   &  15(3)    &  -30(12) \\
24    &   27         &         4          & 15.1   &  20(7)    &  -26(6) \\
32    &   19         &         5          & 11.7   &  26(9)    &  -51(20) \\
40    &   18         &         4          &  9.4   &  155(143) &  -21(6) \\
48    &   13         &         2          &  7.8   &  21(5)    &  -21(5) \\
56    &   12         &         3          &  7.6   &  16(4)    &  -21(7) \\
64    &   11         &         2          &  6.8   &  20(7)    &  -42(18) \\
72    &   11         &         2          &  6.1   &  129(96)  &  -42(13) \\
80    &   12         &         3          &  6.1   &  15(4)    &  -24(4) \\ \hline
\end{tabular}
\end{center}
\caption{
\label{tab:nandR}
The maximum, minimum and average number of steps $n$,
necessary to ``cross the Gribov horizon'' along the direction
$A_{\mu}^b(x)$, as a function of the lattice size $N$. We also show
the ratio $R$ [see Eq.\ (\protect\ref{eq:ratio})], divided by
1000, for the modified gauge fields ${\widehat A}_{\mu}^{(n-1)}(x)
= \tau_{n-1} \; A_{\mu}(x)$ and ${\widehat A}_{\mu}^{(n)}(x) =
\tau_n \; A_{\mu}(x)$, i.e.\ for the configurations immediately
before and after crossing $\partial \Omega$.
}
\end{table}

\section{Simulating the Math}

In order to verify the new bounds presented in the previous section,
we started by considering the third property of the region $\Omega$,
i.e.\ the fact that $\Omega$ is bounded in every direction. To this
end we ``simulate'' the mathematical proof of this property, i.e.\
given a thermalized gauge configuration\footnote{In our
simulations we used 70 thermalized configurations, for the SU(2)
case at $\beta = 2.2$, for lattice volumes $V=16^4$, $24^4, 32^4,
40^4$ and 50 configurations (at the same $\beta$ value) for
lattice volumes $V=48^4, 56^4, 64^4, 72^4, 80^4$.} $A_{\mu}(x)$,
we apply the scale transformations\footnote{With this rescaling we
withdraw the unitarity of the link variables, thus losing the
connection with the usual Monte Carlo simulations. In this sense,
our approach is an informal, free and easy one. Nevertheless, it
gives us useful insights into the properties of the Faddeev-Popov
matrix ${\cal M}$ and of the first Gribov region $\Omega$. Note
also that the rescaled field still respects the gauge condition.}
$ \, {\widehat A}_{\mu}^{(i)}(x) \, = \, \tau_i \; A_{\mu}(x) \, $
such that: {\bf a}) $\tau_0 = 1$, {\bf b}) $\tau_i = \delta \,
\tau_{i-1}$, {\bf c}) $\delta = 1.001$ if $\lambda_1 \geq 5 \,
\times \, 10^{-3}$, {\bf d}) $\delta = 1.0005$ if $\lambda_1 \in
[ 5 \, \times \, 10^{-4}, \, 5 \, \times \, 10^{-3})$ and {\bf e})
$\delta = 1.0001$ if $\lambda_1 < 5 \, \times \, 10^{-4}$, with
$\lambda_1$ evaluated at the step $i-1$. Clearly, after $n$ steps,
the modified gauge field ${\widehat A}_{\mu}^{(n)}(x)$ does not belong
to the region $\Omega$ anymore, i.e.\ the eigenvalue $\lambda_1$ of
${\cal M}[{\widehat A}^{(n)}]$ is negative (while the eigenvalue
$\lambda_2$ is still positive). Results for the number of steps $n$
necessary to ``cross the Gribov horizon'' along the direction
$A_{\mu}^b(x)$ are reported in Table \ref{tab:nandR}. It is
interesting to note that the value of $n$ decreases as the lattice
side $N$ increases, i.e.\ configurations with larger physical volume
are (on average) closer to the boundary $\partial \Omega$.

\begin{figure}[t]
\begin{center}
\vskip -0.6cm
\hskip -5mm
\includegraphics[scale=0.80]{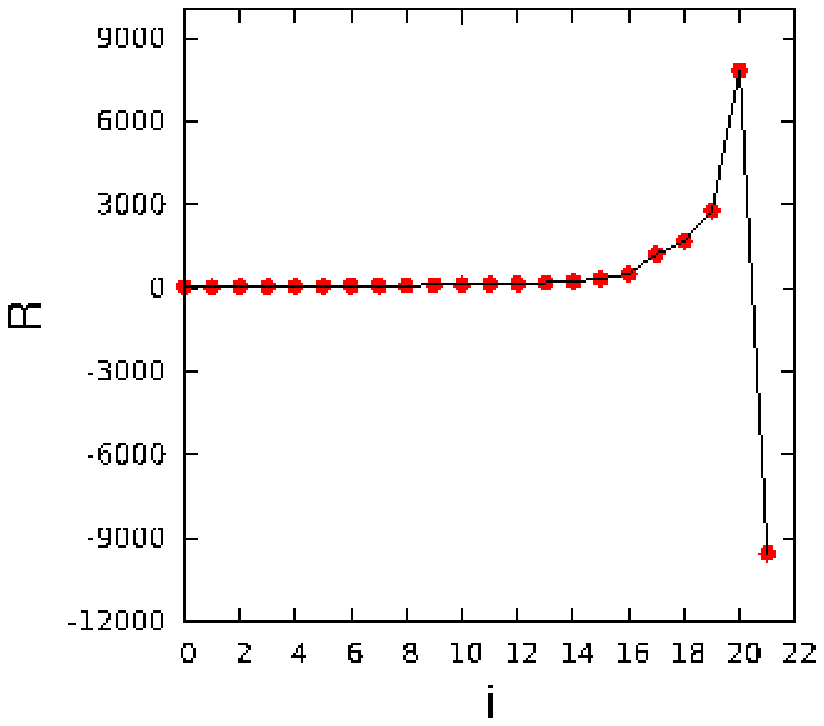}
\hskip 6mm
\includegraphics[scale=0.80]{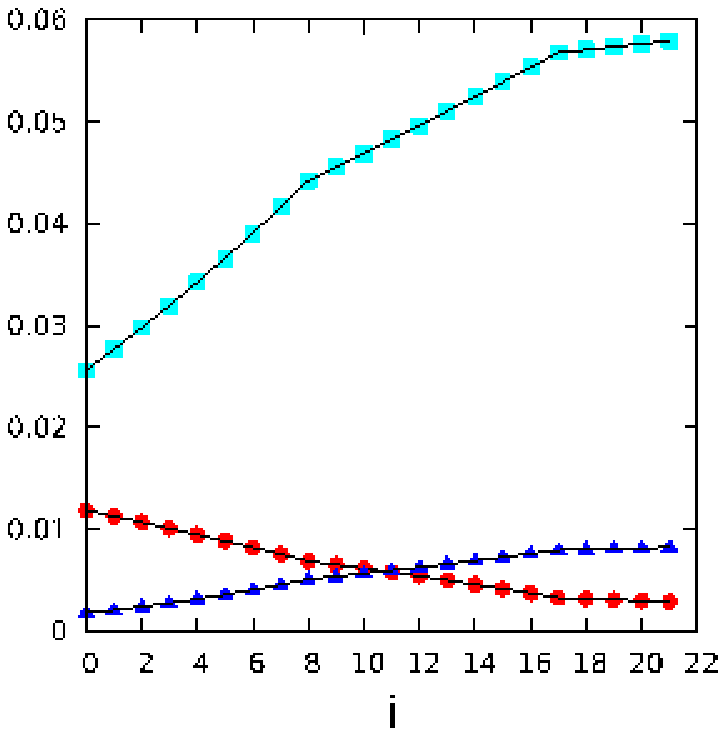}
\vskip 3mm
\caption{
\label{figs:ratios}
Left: plot of the ratio $R$ [see Eq.\ (\protect\ref{eq:ratio})],
as a function of the iteration step $i$, for a typical configuration
and lattice volume $V=16^4$. Right: plot of $\lambda_2$
(\textcolor{red}{\bf full circes}), $| {\cal E}''' \, |$
(\textcolor{cyan}{\bf full squares}) and $ {\cal E}'''' \, $
(\textcolor{blue}{\bf full triangles}) as a function of the
iteration step $i$, for the same configuration considered in
the left plot. 
}
\end{center}
\end{figure}

In the same table we also show the value of the ratio\footnote{Here,
${\cal E}'''$ and ${\cal E}''''$ are the third and the fourth
derivatives of the minimizing functional, defined in Eq.\
(\ref{eq:calE}), evaluated along the direction of the eigenvector
$\psi_1(b,x)$ corresponding to the eigenvalue $\lambda_1$. As shown
in Ref.\ \cite{Cucchieri:1997ns}, this ratio characterizes the
shape of the minimizing functional ${\cal E}$, around the local
minimum considered, when one applies to ${\cal E}$ a fourth-order
Taylor expansion (see in particular Figure 2 of the same
reference).}
\begin{equation}
R \, = \, \frac{({\cal E}''')^2}{\lambda_1 \, {\cal E}''''}
\label{eq:ratio}
\end{equation}
for the rescaled gauge fields ${\widehat A}_{\mu}^{(n-1)}(x) =
\tau_{n-1} \; A_{\mu}(x)$ and ${\widehat A}_{\mu}^{(n)}(x) = 
\tau_n \; A_{\mu}(x)$, i.e.\ for the configurations immediately
before and after crossing the first Gribov horizon $\partial
\Omega$. The same ratio is also shown in Fig.\ \ref{figs:ratios}
(left plot), as a function of the iteration step $i$, for a typical
configuration and for the lattice volume $V=16^4$. For the same
configuration we also show (see Fig.\ \ref{figs:ratios}, right
plot) the dependence of $ \lambda_2 $, $| {\cal E}''' \, |$ and
of $ {\cal E}'''' \, $ on the iteration step $i$. One clearly sees
that these quantities have a slow and continuous dependence on
the factors $\tau_i$. On the other hand, since $\lambda_1$ decreases
as $\tau_i$ increases, we find that the ratio $R$ usually
increases\footnote{However, for a few configurations, we found 
\cite{Cucchieri:2013nja} a very small value for the ratio $R$
for all factors $\tau_i$, i.e.\ also when the configuration
${\widehat A}_{\mu}^{(i)}(x)$ is very close to $\partial \Omega$.
We interpret these configurations as possible candidates to belong
to the common boundary $\partial \Omega \, \cap \, \partial\Lambda$.
Here $\Lambda$ is the fundamental modular region \cite{Zwanziger:1991ac,
Zwanziger:1993dh}, obtained by considering absolute minima of the
minimizing functional ${\cal E}[U;\omega]$ defined in Eq.\
(\ref{eq:calE}).} with $\tau_i$ and that $R_{n-1} \approx - R_n$,
due to the change in sign of $\lambda_1$ as the first Gribov horizon
is crossed (see the fourth and the fifth columns in Table
\ref{tab:nandR} and the left plot in Fig.\ \ref{figs:ratios}). At
the same time one can check (see the right plot in Figure
\ref{figs:ratios}) that the second smallest (non-trivial) eigenvalue
$\lambda_2$ stays positive, i.e.\ the final configuration
${\widehat A}_{\mu}^{(n)}(x) = \tau_n \; A_{\mu}(x)$ belongs to
the second Gribov region \cite{Gribov:1977wm,Zwanziger:1991ac}.

\begin{figure}[t]
\begin{center}
\vspace{-3mm}
\hskip -5mm
\includegraphics[scale=0.80]{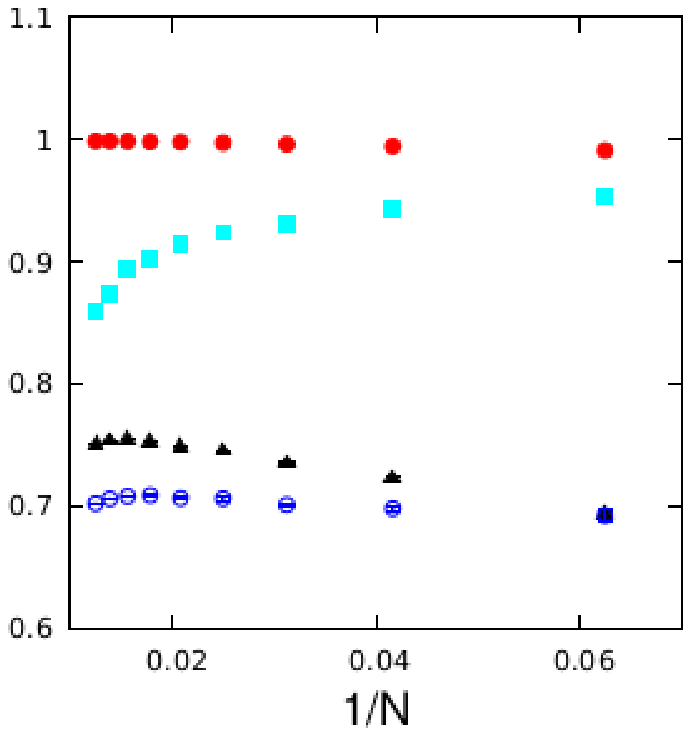}
\hskip 6mm
\includegraphics[scale=0.80]{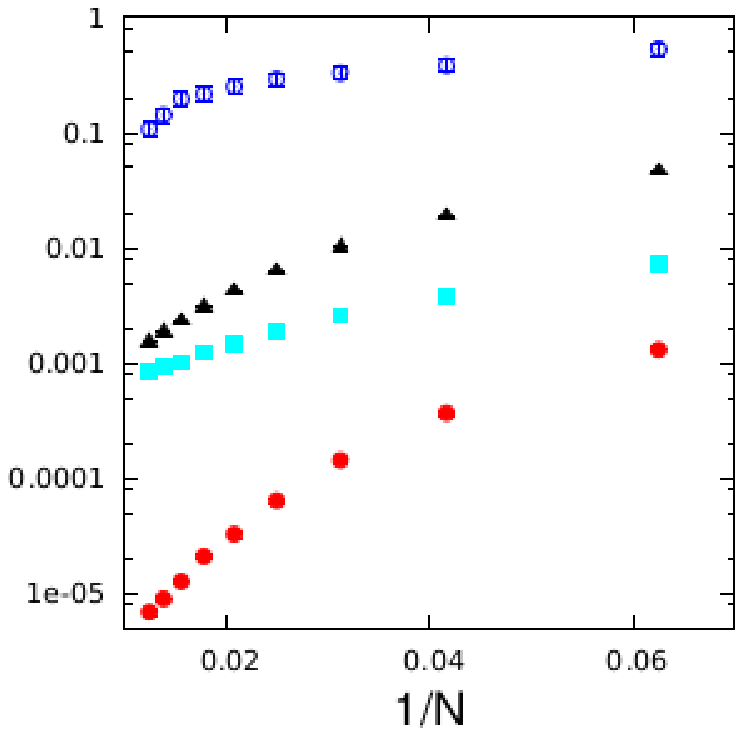}
\vskip 3mm
\caption{\label{fig:ineqnew}
Left: plot of the (normalized) horizon function $h$
(\textcolor{blue}{\bf empty circles}), of the Gribov ghost
form-factor $\sigma(p_{min})$ ({\bf full triangles}), of
the quantity $1 - \lambda_1/p_{min}^2$ (\textcolor{cyan}{\bf
full squares}) and of their upper bound $\rho$ (\textcolor{red}{\bf
full circles}) as a function of the inverse lattice size $1/N$.
Let us note that, in Ref.\ \cite{Capri:2012wx}, it was proven
that $\sigma(0) = h$ to all orders in the gauge coupling.
Right: plot of the inverse of the lower bound
in Eq.\ (\protect\ref{eq:lowerb}) (\textcolor{blue}{\bf empty
circles}), of $1/G(p_{min})$ ({\bf full triangles}), of
$\lambda_1$ (\textcolor{cyan}{\bf full squares}) and of the
quantity $(1 - \rho) \, p^2_{min}$ (\textcolor{red}{\bf
full circles}) as a function of the inverse lattice size
$1/N$.
}
\end{center}
\end{figure}

Once we have found a configuration ${\widehat A}_{\mu}^{(n)} \notin
\Omega$, we can use the definition
\begin{equation}
A_{\mu}'(x) \, = \, \frac{{\widehat A}_{\mu}^{(n-1)}(x) +
                 {\widehat A}_{\mu}^{(n)}(x)}{2}
   \, = \, \frac{\tau_{n-1} + \tau_{n}}{2} \, A_{\mu}(x)
   \, \equiv \, {\widetilde \tau} \, A_{\mu}(x)
\end{equation}
as a candidate for a configuration belonging to the boundary
$\partial \Omega$. This gives us an estimate for the parameter
$\rho = 1/ {\widetilde \tau} < 1$ and allows us to test the
inequalities presented in the previous Section. The numerical
data tell us that most lattice configurations $A$ are
very close to the first Gribov horizon $\partial \Omega$,
i.e.\ one usually finds $\rho \approx 1$ (see Figure
\ref{fig:ineqnew}, left plot). Moreover, the quantity $1-\rho$
goes to zero reasonably fast. On the other hand, the inequality
(\ref{eq:ineqnew2}) is far from being saturated by the lattice data
(see Figure \ref{fig:ineqnew}, right plot), and the situation seems
to become worst in the infinite-volume limit. The same observation
applies (see Fig.\ \ref{fig:ineqnew}, right plot) to the lower bound
in Eq.\ (\ref{eq:lowerb}). Finally, for large lattice size $L$ one
finds $ \, G(p_{min}) \, \ltapprox \, 1/\lambda_1 \, $ (see again
Figure \ref{fig:ineqnew}, right plot). These results are
all consistent with the fact that the eigenvector $\psi_1$ is
very different from the plane waves corresponding to $p^2_{min}$,
which clarifies why the ghost propagator $G(p)$ is not enhanced
in the IR limit. Conversely, configurations producing an
IR-enhanced ghost propagator should almost saturate the new
bound (\ref{eq:ineqnew2}), i.e.\ their eigenvector $\psi_1$ should
have a large projection on at least one of the plane waves
corresponding to $p^2_{min}$. Thus, in the scaling 
solution \cite{von Smekal:1997vx},
nonperturbative effects, such as color confinement, should be driven
by configurations whose FP matrix ${\cal M}$ is ``dominated'' by an
eigenvector $\psi_1$ very similar to the corresponding eigenvector
of ${\cal M} = -\partial^2 \,$, i.e.\ to the eigenvector $\psi_1$
of the free case! This would constitute a very odd situation indeed.
On the contrary, the massive solution of the
DSEs of gluon and ghost propagators \cite{Aguilar:2004sw} is
consistent with the more reasonable hypothesis that the eigenvector
$\psi_1$ is in general very different from a free wave.

\end{document}